\documentclass[aps,prb,preprint,superscriptaddress,showpacs,showkeys,floatfix]{revtex4}

\usepackage{graphicx,color}
\graphicspath{{figs/}}

\begin{document}
\title{Modulation of Thermal Conductivity in Kinked Silicon Nanowires: Phonon Interchanging and Pinching Effects}
\author{Jin-Wu Jiang}
    \altaffiliation{Electronic address: jinwu.jiang@uni-weimar.de}
    \affiliation{Institute of Structural Mechanics, Bauhaus-University Weimar, Marienstr. 15, D-99423 Weimar, Germany}
\author{Nuo~Yang}
    \altaffiliation{Electronic address: imyangnuo@tongji.edu.cn}
    \affiliation{Center for Phononics and Thermal Energy Science, School of Physical Science and Engineering, Tongji University, 200092 Shanghai, China}
\author{Bing-Shen~Wang}
    \affiliation{State Key Laboratory of Semiconductor Superlattice and Microstructure and Institute of Semiconductor, Chinese Academy of Sciences, Beijing 100083, China}
\author{Timon Rabczuk}
    \affiliation{Institute of Structural Mechanics, Bauhaus-University Weimar, Marienstr. 15, D-99423 Weimar, Germany}

\date{\today}
\begin{abstract}
We perform molecular dynamics simulations to investigate the reduction of the thermal conductivity by kinks in silicon nanowires. The reduction percentage can be as high as $70\%$ at room temperature. The temperature dependence of the reduction is also calculated. By calculating phonon polarization vectors, two mechanisms are found to be responsible for the reduced thermal conductivity: (1) the interchanging effect between the longitudinal and transverse phonon modes and (2) the pinching effect, i.e a new type of localization, for the twisting and transverse phonon modes in the kinked silicon nanowires. Our work demonstrates that the phonon interchanging and pinching effects, induced by kinking, are brand new and effective ways in modulating heat transfer in nanowires, which enables the kinked silicon nanowires to be a promising candidate for thermoelectric materials.
\end{abstract}

\maketitle

\section{Introduction}
In 2009, a new type of building block - kinked silicon nanowire (KSiNW) - was synthesized by Tian {\it et al.} in Leiber's group at Harvard University.\cite{TianB} Particularly, they were able to manipulate the arm length of the kink by controlling the growing time. Since then, great experimental efforts have been devoted to investigating the growth mechanism of the kinks in silicon nanowires (SiNWs).\cite{ChenH,YanC,KimJ,PevznerA,ShinN,MusinIR} Besides these experimental works, Schwarz and Tersoff have proposed a theoretical model to interpret the growth mechanism of KSiNWs. In their model, the formation of kinks is due to the interplay of three basic processes: facet growth, droplet statics, and the introduction of new facets.\cite{SchwarzKW,SchwarzKWprl} Stimulated by KSiNWs, several groups have synthesized kinks in other nanowires, such as ${\rm In_{2}O_{3}}$ multi-kinked nanowires,\cite{ShenG} kinked germanium nanowires,\cite{KimJH} germanium–silicon axial heterostructure with kinks,\cite{DayehSA} and kinked ZnO nanowires.\cite{LiS} While existing works for KSiNW mainly concentrate on its growth mechanism, a new experiment from Leiber's group has demonstrated the application of KSiNWs as nanoelectronic bioprobes.\cite{XuL} However, the study of thermal properties like thermal conductivity in KSiNWs is quite limited, although this is important for its application in thermoelectrics.

Recently, the thermal transport of SiNWs has attracted broad interest, because physical understanding of the thermal properties in nano-materials remains unclear.\cite{BalandinAA1998,KhitunA,VolzSG,KhitunA2000,LiD,ZouJ,PokatilovEp,LiangLH,PonomarevaI,YangN2008,MarkussenT,YangN2010,MartinP2009,MartinP2010,LimJ,NikaDL2012,ChenJ,SchellingPK,BodapatiA,HochbaumAI,BoukaiAI,WangJ} Volz and Chen predicted the ultra-low thermal conductivity of SiNWs, which was about two orders of magnitude smaller than the thermal conductivity of bulk Si crystals obtained from molecular dynamics (MD) simulations.\cite{VolzSG} A few years later, Li {\it et al.} showed experimentally the ultra-low thermal conductivity of single SiNW.\cite{LiD} In 1998, the phonon confinement was firstly proposed as a mechanism for the thermal conductivity reduction in low-dimensional structures by Balandin and Wang,\cite{BalandinAA1998} and this idea was worked out specifically to SiNWs and Si-Ge NWs in the following years.\cite{KhitunA,KhitunA2000,ZouJ} Ponomareva {\it et.al} also observed this phonon confinement effect on the thermal conductivity in SiNWs in 2007.\cite{PonomarevaI} Yang {\it et.al} found very low thermal conductivity in the isotope-doped SiNWs.\cite{YangN2008} By introducing surface roughness, it was found that the thermal conductivity in nanowires can be greatly reduced from both theoretical\cite{MartinP2009,MartinP2010} and experimental examinations\cite{LimJ}. For pristine SiNWs, the thermal conductivity is found to be strongly anisotropic.\cite{MarkussenT} Due to the effects of boundary scattering and confinement on lattice thermal conductivity, there is a significant progress on the applications of SiNWs. For example, SiNWs can be used in thermoelectric systems where ultra-low thermal conductivity materials are needed.

In this letter, we perform non-equilibrium MD simulations to study the thermal conductivity in KSiNWs and compare it to that of straight silicon nanowires (SSiNWs). We find that the thermal conductivity of SSiNWs can be reduced up to $70\%$ by kinks at room temperature and the reduction percentage depends on the temperature. After comparatively studying the phonon modes in the SSiNW and the KSiNW, we show that the reduction in the thermal conductivity comes from the kink-induced interchanging and pinching effects on some phonon modes in the KSiNW.

\section{structure and calculation details}
\begin{figure}[htpb]
  \begin{center}
    \scalebox{1.0}[1.0]{\includegraphics[width=8cm]{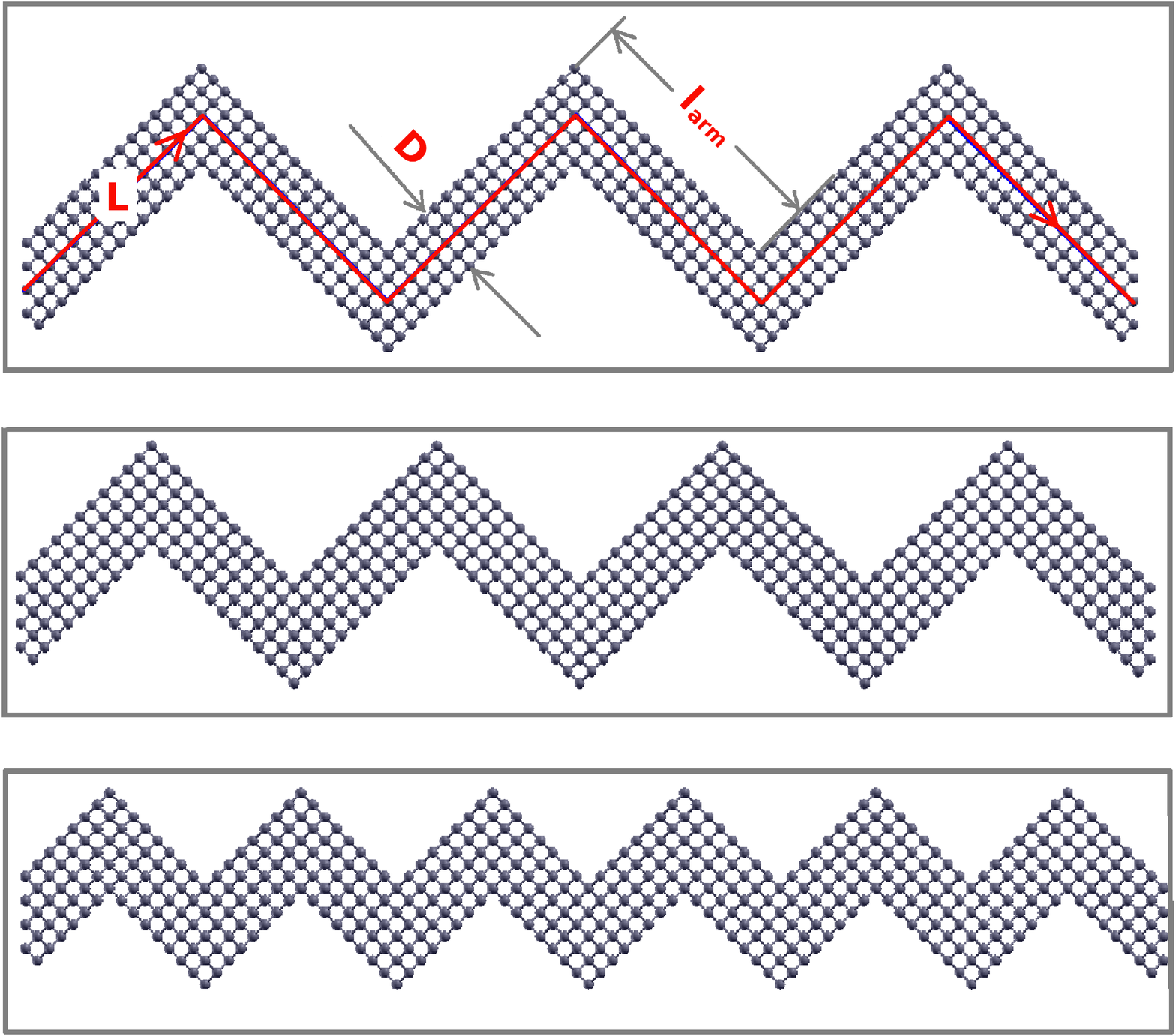}}
  \end{center}
  \caption{Configuration for KSiNWs with diameter $D=11.0~{\AA}$ and axial length $L=190.0~{\AA}$. $l_{\rm arm}$ is the arm length. The number of kinks are 5, 7, and 11 in structures from top to bottom.}
  \label{fig_cfg}
\end{figure}

Fig.~\ref{fig_cfg} displays the configuration of three KSiNWs of 11.0~{\AA} in diameter and 190.0~{\AA} in length along the axial direction. The growth direction changes from one $<110>$ to another $<110>$ direction at the kink. The angle between these two $<110>$ directions is $90^{\circ}$. Fixed boundary condition is applied in the growth direction and free boundary condition is applied in the lateral direction. The force between silicon atoms in the Newton equation is described by the commonly used Tersoff bond order potentials.\cite{Tersoff} The Newton equations are solved by the velocity Verlet algorithm with an integration time step of 1.0 fs. A total simulation time is typically around 2.0 ns and will be sufficiently extended to guarantee the achievement of steady state whenever necessary.

The two regions close to the left/right ends are maintained at high/low temperatures $T_{L/R}=(1\pm\alpha)T$ by N\'ose-Hoover thermostat.\cite{Nose,Hoover} $T$ is the average temperature. $\alpha=0.1$ is adopted in our calculation. The heat energy is pumped into the system through the left temperature-controlled region with thermal current $J_{L}$ and will flow out of the right temperature-controlled region with current $J_{R}$. At steady state, we have $J_{L}=-J_{R}$ as required by the energy conservation law. Using this relation, the thermal current across the system can be obtained by $J=(J_{L}-J_{R})/2$, with $dJ=J_{L}+J_{R}$ as an estimated error.\cite{JiangJW2010jap} The thermal conductivity is obtained from the Fourier law. There is no quantum correction here, as all simulations are performed above room temperature. The quantum correction will become more important for lower temperature. For the lowest temperature in present work, i.e 300~{K}, the quantum correction can be estimated to be around $3.0\%$, following the approach suggested by Maiti, Mahan, and Pantelides.\cite{Maiti}

\section{results and discussion}
\begin{figure}[htpb]
  \begin{center}
    \scalebox{1.0}[1.0]{\includegraphics[width=8cm]{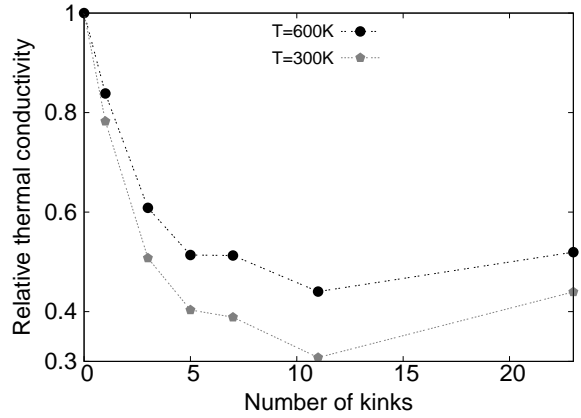}}
  \end{center}
  \caption{Relative thermal conductivity versus number of kinks in KSiNWs of 11.0~{\AA} in diameter and 190.0~{\AA} in length along the axial direction. The value of thermal conductivity for the SSiNW are 7.67 and 5.78~{Wm$^{-1}$K$^{-1}$} at 300 and 600~{K}, respectively.}
  \label{fig_kappa_narm}
\end{figure}

Fig.~\ref{fig_kappa_narm} shows the relative thermal conductivity versus the number of kinks of KSiNWs. The relative thermal conductivity of 1 corresponds to the value of thermal conductivity for the SSiNW, which are 7.9 and 5.8~{Wm$^{-1}$K$^{-1}$} at 300 and 600~{K}, respectively. All systems have the same diameter ($D=11.0$~{\AA}) and the same length ($L=190.0$~{\AA}) along the axial direction. We have observed similar reduction in KSiNWs with other diameter and length. As expected, the reduction of the thermal conductivity is obvious in the kinked structures. The relative thermal conductivity decreases to $44\%$ and $31\%$ as the kink number increases from 0 to 11 at temperature 300K and 600 K, respectively. However, the relative thermal conductivity increases when keep increasing the kink number. In our simulation structures, the arm length $l_{\rm arm}$ (see Fig.~\ref{fig_cfg}) is shortened as the kink number increases, which may be convenient for phonon transport. Similarly, there is also a minimum thermal conductivity when increasing the interface number in superlattice structures.\cite{Simkin,YangN2008,JiangJW2011} For ballistic heat transport, the minimum thermal conductivity results from the competition between the number and the localization properties of the confined phonon modes.\cite{JiangJW2011} For diffusive heat transport, the minimum thermal conductivity is attributed to the interplay between the wave and the particle nature of the phonon quasi-particles.\cite{Simkin}

\begin{figure}[htpb]
  \begin{center}
    \scalebox{1.0}[1.0]{\includegraphics[width=8cm]{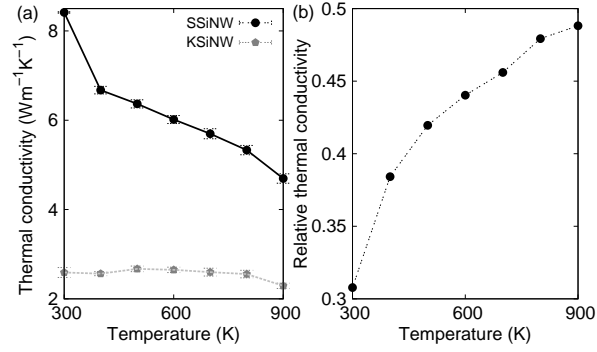}}
  \end{center}
  \caption{Thermal conductivity versus MD simulation temperature. (a) Thermal conductivity in SSiNW (dots) and KSiNW with 11 kinks (pentagons). (b) The ratio of thermal conductivity in KSiNW to that in SSiNW. The lines are guides to the eye.}
  \label{fig_kappa_temperature}
\end{figure}

As shown in Fig.~\ref{fig_kappa_narm}, the reduction of the thermal conductivity also depends on the temperature. To investigate the temperature effect further, we show the temperature dependence of the thermal conductivity of both SSiNWs and KSiNWs in Fig.~\ref{fig_kappa_temperature}~(a). The thermal conductivity in SSiNW decreases as the temperature increases because phonon-phonon scattering dominate the heat transport at high temperature. However, the thermal conductivity in KSiNW is insensitive to the temperature. Fig.~\ref{fig_kappa_temperature}~(b) shows the thermal conductivity ratio of KSiNW to SSiNW based on the data in Fig.~\ref{fig_kappa_temperature}~(a). The relative thermal conductivity increases monotonically with increasing temperature, because the thermal conductivity in SSiNWs is largely deduced at high temperature and the thermal conductivity in KSiNWs is temperature insensitive. There are less phonon-phonon scatterings and a higher thermal conductivity in SSiNWs at low temperature, where the phonon pinching is more effective in confining the heat transfer.

\begin{figure}[htpb]
  \begin{center}
    \scalebox{1.0}[1.0]{\includegraphics[width=8cm]{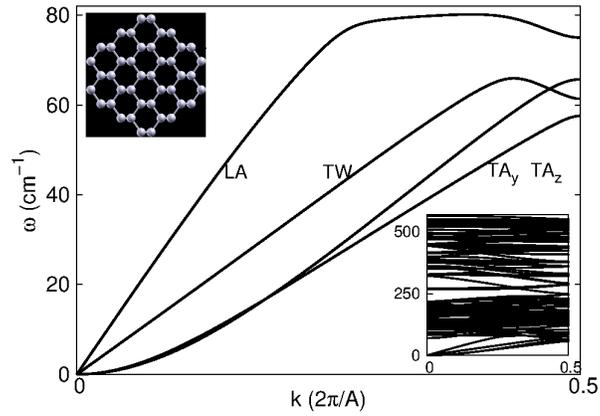}}
  \end{center}
  \caption{Phonon dispersion of the four low-frequency acoustic modes: LA, TW, and two TA modes in SSiNW of 11.0~{\AA} in diameter and period length $A=33.0$~{\AA}. Left top inset gives the cross-section of the nanowire. Right bottom inset shows the full phonon dispersion in the first Broullin zone.}
  \label{fig_phonon_110_d_11}
\end{figure}

To reveal the microscopic origin for the reduction of the thermal conductivity in the KSiNW, we comparatively study the phonon modes in the SSiNW and KSiNW, which are the only energy carrier in the non-metal solids. In Fig.~\ref{fig_phonon_110_d_11}, we show the dispersion of all four acoustic branches of SSiNW\cite{JiangJW2008}, which are one longitudinal acoustic (LA), one twisting (TW), and two transverse acoustic (TA) modes. The full phonon dispersion of SSiNW is shown in the right bottom inset of Fig.~\ref{fig_phonon_110_d_11}.

To show the pinching effect in modes, we calculate the eigen modes of the four acoustic branches in the SSiNW with length $L=190~{\AA}$ and a KSiNW with one kink and $95~{\AA}$ in arm length. The diameter is kept the same for all structures. Both ends of the nanowire are fixed in this calculation, which is consistent with the fixed boundary condition applied in the above thermal transport simulation. The phonon modes (frequencies and polarization vectors) are obtained by solving the eigenvalue problem of the dynamical matrix derived from the Tersoff potential.\cite{Tersoff} The dynamical matrix is obtained by $K_{ij}=\partial^{2}V/\partial x_{i}\partial x_{j}$, where $V$ is the Tersoff potential and $x_{i}$ is the position of the i-th degree of freedom. This formula is realized numerically by calculating the energy change after a small displacement of the i-th and j-th degrees of freedom. The obtained phonon modes from this calculation for the finite system are related to the traveling phonons in Fig.~\ref{fig_phonon_110_d_11} for a bulk structure (with periodic boundary condition applying on a small unit cell). The former is essentially a special solution of the latter under particular boundary conditions (fix boundary conditions here), so the phonon modes in the finite system can be decomposed into some traveling waves which will change into the corresponding reflection and tunneling waves at the kinks. As a result, the phonon modes in the finite system provide intrinsic information for the thermal transport.

\begin{figure}[htpb]
  \begin{center}
    \scalebox{1.0}[1.0]{\includegraphics[width=8cm]{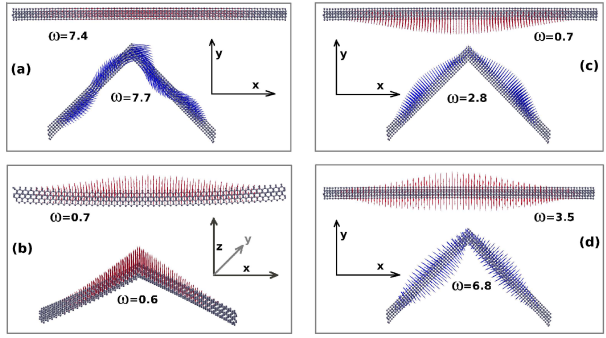}}
  \end{center}
  \caption{(Color online) Comparison between the four low-frequency acoustic modes in the SSiNW with length $L=190~{\AA}$ and the KSiNW with arm length $l_{\rm arm}=95~{\AA}$. Arrows in the figure represent the polarization vector of the phonon mode. Red (blue) color indicates good (poor) heat transport capability for the mode. (a) LA mode. (b) TA mode vibrating in $z$ direction. (c) TA mode vibrating in $y$ direction. (d) TW mode. Numbers in the figure are the corresponding frequency in unit of cm$^{-1}$. Note the pinching effect for the two modes in the KSiNW in (c) and (d).}
  \label{fig_u}
\end{figure}

In the following, the eigen modes in the SSiNW are compared to those in the KSiNW. Four modes with the lowest frequency in each acoustic branch in SSiNWs are selected and presented in Fig.~\ref{fig_u}. The polarization vectors of the eigen-modes are represented by arrows whose length corresponds to the amplitudes. Red (blue) arrows indicate a mode with good (poor) heat transport capability; i.e the phonon is extended (pinched). To analyze the correspondence between phonon modes from both structures in Fig.~\ref{fig_u}, we have examined the polarization vectors of the first thirty lowest-frequency phonon modes in the SSiNW and the KSiNW.

Fig.~\ref{fig_u}~(a) shows the polarization vectors of a LA mode (along x direction) with $\omega=7.4$~{cm$^{-1}$} in the SSiNW. In the KSiNW, the polarization vector of a LA mode with $\omega=7.7$~{cm$^{-1}$} is not confined in x direction. When the phonon transfers along the KSiNW from one end to the other, the LA mode and TA$_{y}$ mode interchange into each other, which leads to the lose of the coherence of the phonon mode and eventually reduces its heat transport capability. This deduction effect is also observed in the folded graphene nanoribbon structure.\cite{YangN2012} The frequency of the LA mode in the KSiNW is only slightly higher than that in the SSiNW due to the kink effect. We note that there is a mirror reflection symmetry locates at the middle of the kin angle in the KSiNW. Due to this mirror symmetry, there is another LA mode in the KSiNW with closer frequency (8.0~cm$^{-1}$).

In Fig.~\ref{fig_u}~(b), we compare a ${\rm TA_{z}}$ mode in the SSiNW and the KSiNW, whose polarization vectors are all along the $z$-direction. The ${\rm TA_{z}}$ mode is almost not affected by the kink interface, owing to its special vibrating morphology which is perpendicular to the kink ($xy$ plane). There is barely a difference in the frequency and vibration between the ${\rm TA_{z}}$ modes in SSiNW and that in KSiNW. Hence, the ${\rm TA_{z}}$ modes are good at heat delivery in both SSiNWs and KSiNWs.

In the following, the phonon pinching is shown obviously in ${\rm TA_{y}}$ and TW modes of KSiNWs. As shown in Fig.~\ref{fig_u}~(c), there is a critical difference between the ${\rm TA_{y}}$ mode in SSiNW and that in KSiNW. The ${\rm TA_{y}}$ mode with $\omega=0.7$~{cm$^{-1}$} in the SSiNW is a quite extended mode. While in the KSiNW, it is interesting that vibrating amplitudes of the ${\rm TA_{y}}$ mode ($\omega=2.8$~{cm$^{-1}$}) are nearly zero at the kink. In other words, the polarization vector of this phonon mode is pinched-off by the kink. The polarization vectors of the ${\rm TA_{y}}$ modes are symmetric about the mirror plane at the kink. There is another ${\rm TA_{y}}$ with a closer frequency (2.0~cm$^{-1}$) in the KSiNW, whose polarization vectors are anti-symmetric about the mirror plane, and the phonon pinching effect is also found there.

The pinching-off picture can be validated by the relationship between frequencies of corresponding modes in SSiNWs and KSiNWs. The frequency of the mode in the KSiNW (2.8~cm$^{-1}$) is four times of the frequency in the SSiNW (0.7~cm$^{-1}$). The arm length of the KSiNW is half of the length of the SSiNW. As shown in Fig.~\ref{fig_phonon_110_d_11}, the frequency of the ${\rm TA_{y}}$ mode is proportional to the square of the wave vector, $\omega=\beta k^{2}=\beta (2\pi/L)^{2}$, due to the flexure nature of the quasi-one-dimensional configuration of the nanowire,\cite{JiangJW2011prb} $\beta$ is a coefficient determined by the Tersoff potential. If a mode in SSiNW is pinched-off from the middle, the transverse mode confined in the left or right arm is degenerated, whose frequencies are $\beta (\pi/L)^{2}$, i.e $4\omega$. This is the origin of the pinched-off transverse mode in the KSiNW. Similar to the ${\rm TA_{y}}$ modes in the KSiNW, the pinching effect is observed in the TW modes shown in Fig.~\ref{fig_u}~(d). Since the frequency for the TW mode is inversely proportional to the length of the nanowire (Fig.~\ref{fig_phonon_110_d_11}), the frequency of the TW modes in the KSiNW is almost twice the frequency in the SSiNW.

\begin{figure}[htpb]
  \begin{center}
    \scalebox{1.0}[1.0]{\includegraphics[width=8cm]{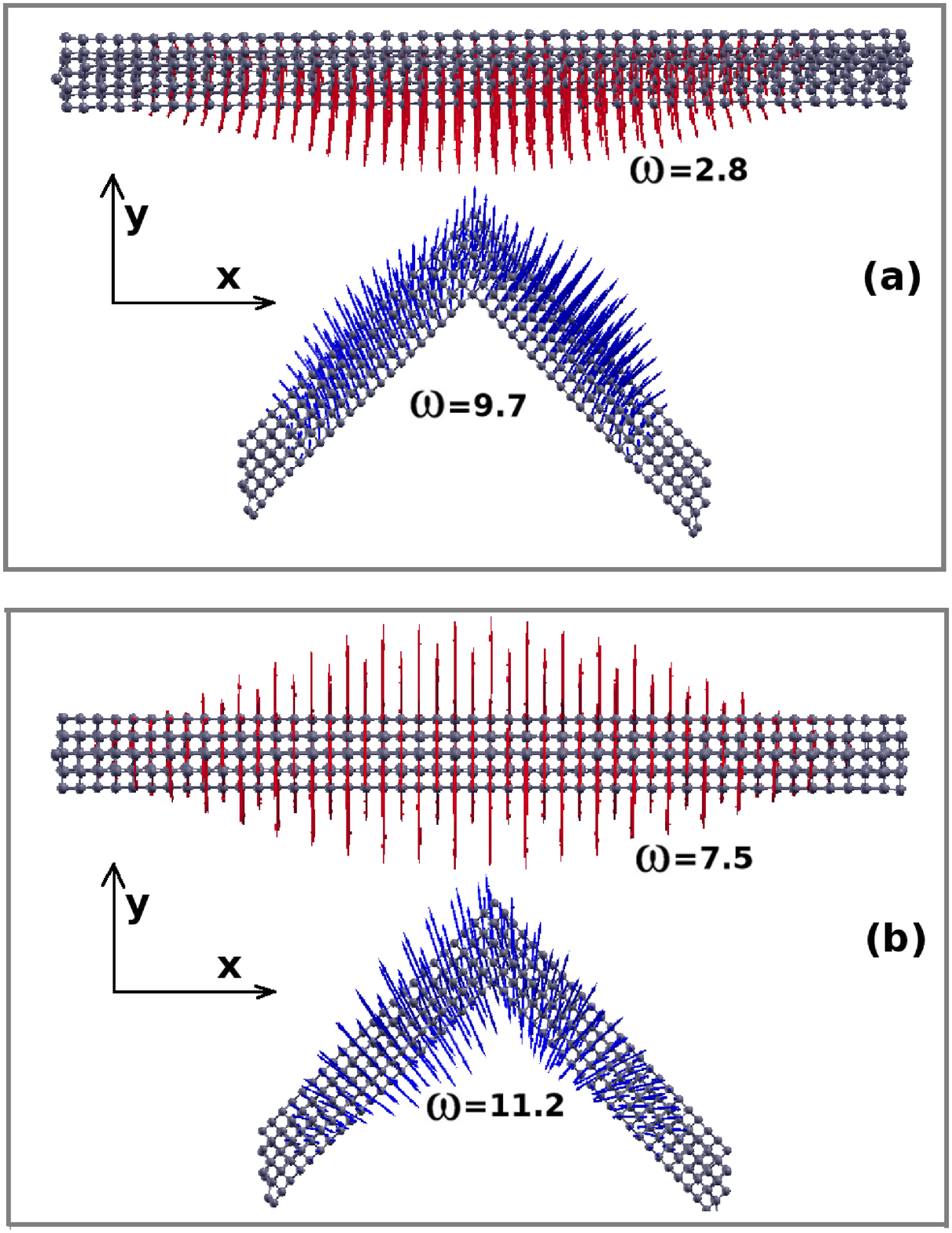}}
  \end{center}
  \caption{(Color online) Phonon pinching effect in the SSiNW with length $L=95~{\AA}$ and the KSiNW with arm length $l_{\rm arm}=47.5~{\AA}$ for (a) TA mode vibrating in $y$ direction, and (b) TW mode. Pinching effect here is weaker than the two corresponding modes in the KSiNW of longer arm length shown in Fig.~\ref{fig_u}~(c) and (d).}
  \label{fig_u_short}
\end{figure}

As a result of the pinching effect, the heat energy carried by the ${\rm TA_{y}}$ and the TW modes is difficult to be transported from the left arm to the right arm of the KSiNW. The pinched phonon mode is a new type of localization mode. This is the main reason for the reduction of the thermal conductivity by kinks in the KSiNW. The dependence of the reduction in the thermal conductivity of a KSiNW with fixed axial length ($L$) on kinks shown in Fig.~\ref{fig_kappa_narm} can be understood as follows: on the one hand, there are more pinching effects in KSiNW with more kinks, so the thermal conductivity becomes smaller with increasing kink number; on the other hand, the arm length is shorter for KSiNW with more kinks, and the pinching effect is weaker in KSiNW of shorter arm length as shown in Fig.~\ref{fig_u_short}. The arm length for the KSiNW studied in Fig.~\ref{fig_u_short} is $l_{\rm arm}=47.5~{\AA}$. We make a comparison between the frequencies in the SSiNW ($\omega_{\rm S}$) and the KSiNW ($\omega_{\rm K}$). It is obvious that $\omega_{\rm K}^{\rm TA}<4\omega_{\rm S}^{\rm TA}$ and $\omega_{\rm K}^{\rm TW}<2\omega_{\rm S}^{\rm TW}$, for the ${\rm TA_{y}}$ and the TW modes, respectively. This provides a direct evidence for the weaker pinching effect in KSiNW with shorter arm length, because it is expected that $\omega_{\rm K}^{\rm TA}\approx4\omega_{\rm S}^{\rm TA}$ and $\omega_{\rm K}^{\rm TW}\approx2\omega_{\rm S}^{\rm TW}$ in case of a strong pinching effect such as that in Fig.~\ref{fig_u}. Related to this weaker pinching effect, the vibrating amplitude of the atoms in the kink region starts to arise, which induces a tunneling phenomenon of this mode at the kink. This tunneling effect facilitates the heat delivery across the kink, thus enhancing the thermal conductivity of the KSiNW. The interplay between the pinching enhancement effect (due to increasing kink number) and the tunneling enhancement effect (due to the shortening of arm length) yields the minimum thermal conductivity in Fig.~\ref{fig_kappa_narm}.

We would like to stress that lower thermal conductivity in nanowires is essential for their application in thermoelectric materials. There are some other methods to reduce the thermal conductivity, besides the phonon pinching effect discussed in present work. For example, Nika {\it et.al} demonstrated the importance of the phonon filtering effect on the thermal conductivity in one-dimensional superlattice nanowires.\cite{NikaDL2011} They found very low thermal conductivity, because significant number of phonon modes are filtered out from thermal transport. The thermal conductivity can be further manipulated by combing the phonon pinching and other effects (eg. phonon filtering) in nanowires.

\section{conclustion}
We have performed MD simulation to study the heat transport in the KSiNW. We observe that the thermal conductivity in the KSiNW is insensitive to the temperature, since the heat transport in KSiNW is mainly limited by the phonon scattering at the kink interface. We found that the thermal conductivity can be reduced by kinks in the KSiNW for $70\%$ compared to that of the SSiNW at room temperature and the reduction can be further increased at lower temperature. By comparatively investigating the phonon modes in SSiNW and KSiNW, we disclose the underlying mechanism for the reduction of the thermal conductivity to be the kink-induced interchanging effect between LA and ${\rm TA_{y}}$ modes and the pinching effect on ${\rm TA_{y}}$ and the TW modes in the KSiNW. Both interchanging and pinching effects greatly suppress the transport ability of these phonon modes, leading to the reduction of the thermal conductivity.

\textbf{Acknowledgements} We thank H. S. Park at Boston University for useful discussions. The work is supported in part by the Grant Research Foundation (DFG) Germany. N. Y. acknowledges the support by the National Natural Science Foundation of China (Grant No. 11204216) and the startup fund from Tongji University, China.

\pagebreak

\begin{thebibliography}{}
\bibitem{TianB} B. Tian, P. Xie, T. J. Kempa, D. C. Bell, and C. M. Lieber, Nature Nanotech. \textbf{4}, 824 (2009).

\bibitem{ChenH} H. Chen, H. Wang, X.-H. Zhang, C.-S. Lee, and S.-T. Lee, Nano. Lett. \textbf{10}, 864 (2010).

\bibitem{YanC} C. Yan, N. Singh, and P. S. Lee, ACS. Nano. \textbf{4}, 5350 (2010).

\bibitem{KimJ} J. Kim, Y. H. Kim, S.-H. Choi, and W. Lee, ACS. Nano. \textbf{5}, 5242 (2011).

\bibitem{PevznerA} A. Pevzner, Y. Engel, R. Elnathan, A. Tsukernik, Z. Barkay, and F. Patolsky, Nano. Lett. \textbf{1}, 7 (2012).

\bibitem{ShinN} N. Shin and M. A. Filler, Nano. Lett. \textbf{12}, 2865 (2012).

\bibitem{MusinIR} I. R. Musin and M. A. Filler, Nano. Lett. \textbf{12}, 3363 (2012).

\bibitem{SchwarzKW} K. W. Schwarz and J. Tersoff, Nano. Lett. \textbf{11}, 316 (2011).

\bibitem{SchwarzKWprl} K. W. Schwarz, J. Tersoff, S. Kodambaka, Y.-C. Chou, and F. M. Ross, Phys. Rev. Lett. \textbf{107}, 265502 (2011).

\bibitem{ShenG} G. Shen, B. Liang, X. Wang, P.-C. Chen, and C. Zhou, ACS. Nano. \textbf{5}, 2155 (2010).

\bibitem{KimJH} J. H. Kim, S. R. Moon, Y. Kim, Z. Gang Chen, J. Zou, D. Y. Choi, H. J. Joyce, Q. Gao, H. H. Tan, and C. Jagadish, Nanotechnology \textbf{23}, 115603 (2012).

\bibitem{DayehSA} S. A. Dayeh, J. Wang, N. Li, J. Yu Huang, A. V. Gin, and S. T. Picraux, Nano. Lett. \textbf{11}, 4200 (2011).

\bibitem{LiS} S. Li, X. Zhang, L. Zhang, and M. Gao, Nanotechnology \textbf{21}, 435602 (2010).

\bibitem{XuL} L. Xu, Z. Jiang, Q. Qing, L. Mai, Q. Zhang, and C. M. Lieber, DOI: 10.1021/nl304435z, Nano. Lett. (ASAP, 2013).

\bibitem{BalandinAA1998} A. A. Balandin and K. L. Wang, Phys. Rev. B \textbf{58}, 1544 (1998)

\bibitem{KhitunA} A. Khitun, A. A. Balandin, and K. L. Wang, J. Superlattices and Microstructures \textbf{26}, 181 (1999).

\bibitem{VolzSG} S. G. Volz and G. Chen, Appl. Phys. Lett. \textbf{75}, 2056 (1999).

\bibitem{KhitunA2000} A. Khitun, A. A. Balandin, K. L. Wang, and G. Chen, Physica E \textbf{8}, 13 (2000).

\bibitem{ZouJ} J. Zou and A. A. Balandin, J. Appl. Phys. \textbf{89}, 2932 (2001).

\bibitem{SchellingPK} P. K. Schelling, S. R. Phillpot, and Pl Keblinski, Phys. Rev. B \textbf{65}, 144306 (2002).

\bibitem{LiD} D. Li, Y. Wu, P. Kim, L. Shi, P. Yang, and A. Majumdar, Appl. Phys. Lett. \textbf{83}, 2934 (2003).

\bibitem{PokatilovEp} E. P. Pokatilov, D. L. Nika and A. A. Balandin, Phys. Rev. B \textbf{72}, 113311 (2005).

\bibitem{LiangLH} L. H. Liang and B. Li, Phys. Rev. B \textbf{73}, 153303 (2006).

\bibitem{BodapatiA} A. Bodapati, P. K. Schelling, S. R. Phillpot, and P. Keblinski, Phys. Rev. B \textbf{74} (2006).

\bibitem{PonomarevaI} I. Ponomareva, D. Srivastava, and M. Menon, Nano. Lett. \textbf{7}, 1155 (2007).

\bibitem{WangJ} J. Wang and J.-S. Wang, Appl. Phys. Lett. \textbf{90}, 241908 (2007).

\bibitem{YangN2008} N. Yang, G. Zhang, and B. Li, Nano. Lett. \textbf{8}, 276 (2008).

\bibitem{MarkussenT} T. Markussen, A.-P. Jauho, and M. Brandbyge, Nano. Lett. \textbf{8}, 3771 (2008).

\bibitem{HochbaumAI} A. I. Hochbaum, R. Chen, R. D. Delgado, W. Liang, E. C. Garnett, M. Najarian, A. Majumdar, and P. Yang, Nature \textbf{451}, 163 (2008).

\bibitem{BoukaiAI} A. I. Boukai, Y. Bunimovich, J. Tahir-Kheli, J.-K. Yu, W. A. Goddard, and J. R. Heath, Nature \textbf{451}, 168 (2008).

\bibitem{MartinP2009} P. Martin, Z. Aksamija, E. Pop, and U. Ravaioli, Phys. Rev. Lett. \textbf{102}, 125503 (2009).

\bibitem{MartinP2010} P. Martin, Z. Aksamija, E. Pop, and U. Ravaioli, Nano. Lett. \textbf{10}, 1120 (2010).

\bibitem{YangN2010} N. Yang, G. Zhang, and B. Li, Nano Today, \textbf{5}, 85 (2010).

\bibitem{ChenJ} J. Chen, G. Zhang, and B. Li, Nano. Lett. \textbf{10}, 3978 (2010).

\bibitem{NikaDL2012} D. L. Nika, A. I. Cocemasov, C. I. Isacova, A. A . Balandin, V. M. Fomin and O. G. Schmidt, Phys. Rev. B \textbf{85}, 205439 (2012).

\bibitem{LimJ} J. Lim, K. Hippalgaonkar, S. C. Andrews, A. Majumdar, and P. Yang, Nano. Lett. \textbf{12}, 2475 (2012).

\bibitem{Tersoff} J. Tersoff, Phys. Rev. B \textbf{38}, 9902 (1988).

\bibitem{Nose} S. N\'ose, J. Chem. Phys. \textbf{81}, 511 (1984).

\bibitem{Hoover} W. G. Hoover, Phys. Rev. A, \textbf{31}, 1695 (1985).

\bibitem{JiangJW2010jap} J.-W. Jiang, J. Lan, J.-S. Wang, and B. Li, J. Appl. Phys. \textbf{107}, 054314 (2010).

\bibitem{Maiti} A. Maiti, G. D. Mahan, and S. T. Pantelides, Solid State Commun. \textbf{102}, 517 (1997).

\bibitem{JiangJW2011} J.-W. Jiang, J.-S. Wang, and B.-S. Wang, Appl. Phys. Lett. \textbf{99}, 043109 (2011).

\bibitem{Simkin} M. V. Simkin and G. D. Mahan, Phys. Rev. Lett. \textbf{84}, 927 (2000).

\bibitem{JiangJW2008} J.-W. Jiang, H. Tang, B.-S. Wang, and Z.-B. Su, J. Phys.: Condens. Matter \textbf{20}, 045228 (2008).

\bibitem{YangN2012} N. Yang, X. Ni, J.-W. Jiang, and B. Li, Appl. Phys. Lett. \textbf{100}, 093107 (2012).

\bibitem{JiangJW2011prb} J.-W. Jiang, B.-S. Wang, and J.-S. Wang, Phys. Rev. B \textbf{83}, 235432 (2011).

\bibitem{NikaDL2011} D. L. Nika, E. P. Pokatilov, A. A. Balandin, V. M. Fomin, A. Rastelli, and O. G. Schmidt, Phys. Rev. B \textbf{84}, 165415 (2011).

\end{thebibliography}
\end{document}